\documentclass[reprint,nofootinbib, amsmath,amssymb,aps,onecolumn]{revtex4-2}

\usepackage{amssymb,amsmath,amsfonts,amsbsy,graphicx}
\usepackage{color}
\usepackage{psfrag}
\usepackage{stackrel}

\usepackage{csquotes}

\newcommand\be{\begin{equation}}
\newcommand\ee{\end{equation}}
\newcommand\ba{\begin{eqnarray}}
\newcommand\ea{\end{eqnarray}}\newcommand\eq{\begin{equation}}           
\newcommand\en{\end{equation}}

 \newcommand\epm{$e^{\pm}$ }
 
 \newcount\colveccount
\newcommand*\colvec[1]{
        \global\colveccount#1
        \begin{pmatrix}
        \colvecnext
}
\def\colvecnext#1{
        #1
        \global\advance\colveccount-1
        \ifnum\colveccount>0
                \\
                \expandafter\colvecnext
        \else
                \end{pmatrix}
        \fi
}
\input epsf

\usepackage{ulem}
\usepackage{amssymb}
\usepackage{amsmath}
\usepackage{bm}
\usepackage{graphicx}
\usepackage{color}
\usepackage{subfig}
\usepackage{caption}

\def\gsim{\;\rlap{\lower 2.5pt
 \hbox{$\sim$}}\raise 1.5pt\hbox{$>$}\;}
\def\lsim{\;\rlap{\lower 2.5pt
 \hbox{$\sim$}}\raise 1.5pt\hbox{$<$}\;}

\usepackage{graphicx}
\usepackage{url}
\usepackage{amssymb,amsmath}
\usepackage{amsbsy}

\newcommand{\mnras}{MNRAS}

\usepackage{graphicx}
\usepackage{url}
\usepackage{amssymb,amsmath}
\usepackage{amsbsy}

\begin{document}
\title{
 Radio bounds on the mixed dark matter scenarios of primordial black holes and WIMPs  
}
 \author{Kenji Kadota$^{1,2}$ and Hiroyuki Tashiro$^{3}$ \\
{\small  $^1$School of Fundamental Physics and Mathematical Sciences,
  Hangzhou Institute for Advanced Study,\\
  University of Chinese Academy of Sciences (HIAS-UCAS), Hangzhou 310024, China\\
   $^2$International Centre for Theoretical Physics Asia-Pacific (ICTP-AP), Beijing/Hangzhou, China\\
   $^3$Department of Physics and Astrophysics, Nagoya University, Nagoya 464-8602, Japan
}
}

\begin{abstract}
  We study the synchrotron radio emission in the mixed dark matter scenarios consisting of the primordial black holes (PBHs) and the self-annihilating WIMPs (weakly interacting massive particles). The WIMPs can form the ultracompact minihalos around PBHs and the annihilation enhancement from these dense halos can lead to the efficient synchrotron radiation at the radio frequency in the presence of galactic magnetic fields.
  The upper bound of PBH fraction with respect to the total dark matter abundance is of order $10^{-8}\sim 10^{-5}$ depending on the electroweak scale WIMP mass ($m_{\chi}=10\sim 1000$ GeV) and the WIMP annihilation channel (e.g. a hadronic $\chi \chi \rightarrow b \bar{b}$ or a leptonic $\chi \chi \rightarrow e^+ e^-$ channel). The PBH contribution to the total dark matter abundance is hence negligible when the other component of dark matter is composed of the conventional electroweak scale WIMPs.
\end{abstract}

\maketitle   

\setcounter{footnote}{0} 
\setcounter{page}{1}\setcounter{section}{0} \setcounter{subsection}{0}
\setcounter{subsubsection}{0}

\section{introduction}
The nature of dark matter (DM) still remains elusive despite the convincing evidence for its existence from the astrophysical observations. Commonly discussed DM candidates include the weakly interacting massive particles (WIMPs) motivated from supersymmetry and the axion or more generally axion like particles motivated from pseudo-Nambu-Goldstone bosons arising from the spontaneous symmetry
breaking in the early Universe \cite{Jungman:1995df,Bertone:2004pz,Peccei:1977hh,Weinberg:1977ma,Wilczek:1977pj,Preskill:1982cy,Abbott:1982af,Dine:1982ah}. Moreover the DM do not need to be fundamental particles and the detection of gravitational waves \cite{Abbott:2016blz,TheLIGOScientific:2016pea,LIGOScientific:2018jsj} revived the interests in the primordial black hole (PBH) DM. While the parameter space for the PBH to account for the whole DM of the Universe is being narrowed due to the observation data such as those from the gravitational lensing, PBHs being a partial DM component still remains an intriguing possibility \cite{Carr:2020xqk,Green:2020jor,Carr:2020gox}. In particular we consider the scenarios where the DM is composed of PBHs and WIMPs and study the radio observation bounds on the allowed PBH abundance. Most of the previous literature studying such PBH-WIMP mixed DM scenarios focused on the gamma ray and more recently on the CMB and 21cm bounds \cite{Scott:2009tu,Gondolo:1999ef,Lacki:2010zf,Boucenna:2017ghj,Adamek:2019gns,Eroshenko:2016yve,Carr:2020mqm,Cai:2020fnq,Delos:2018ueo,Kohri:2014lza,Bertone:2019vsk,Ando:2015qda,Hertzberg:2020kpm,Yang:2020zcu,Zhang:2010cj,Tashiro:2021xnj,Yang:2020zcu}.
While the synchrotron radio emission bounds on the WIMP annihilation have been actively discussed in the literature, little attention has been given to the synchrotron radiation in the context of PBH-WIMP mixed DM scenarios which is the focus of our paper \cite{Cirelli:2016mrc,Mambrini:2012ue, Borriello:2008gy, Storm:2016bfw,Colafrancesco:2006he,Bertone:2008xr,Regis:2008ij,Paul:2018njd}. The radio observations can give complimentary to and in some cases tighter bounds on the DM properties than those from gamma rays \cite{Storm:2016bfw,Colafrancesco:2006he,Bertone:2008xr,Regis:2008ij,Paul:2018njd}. There also have been the works discussing the radio signals from the evaporation of PBHs via Hawking radiation for which the typical PBH mass range explored is of order $M_{PBH}\sim 10^{15}-10^{17}$g \cite{Dutta:2020lqc, Mittal:2021dpe,Poulter:2019ooo,Mukhopadhyay:2021puu,Chan:2020zry}.
Our radio signal study explores the totally different PBH mass range of order a solar mass ($1 M_{\odot}\sim 2 \times 10^{33}$g).

Because different wavelength observations can have different systematics and sensitivities to the potential signals, it would be of great interest to check if the multi-wavelength observations support or disprove each other's analysis for the further scrutiny of potential DM signals. The DM annihilation can produce the particles covering a wide range of energy spectra.  
In particular, we are interested in the synchrotron radio emission from the energetic electrons/positrons in the presence of the galactic magnetic fields.

We study the synchrotron radio emission from the WIMP halos around PBHs.
The WIMPs can accrete to a PBH to form the ultracompact minihalo (UCMH) around it. The UCMH can possess a steep density profile and the WIMP annihilation which is proportional to the WIMP density squared is expected to be enhanced.  
Among the particles produced from such enhanced WIMP annihilation, the relativistic electrons (or positrons) (\epm can be the WIMP annihilation final state or more commonly can come from the hadronization and cascade decays of final states) are of particular interest in our study because their synchrotron radiation lies in the observable radio frequency. The synchrotron radiation can be indeed the main source of energy loss for electrons when they have the energy of order a few GeV and above typical for the weak scale mass WIMPs. This is especially the case in the Galactic center where the magnetic field can be bigger than that corresponding to the equivalent CMB energy density ($B \gtrsim 3.25 (1+z)^2 \mu G$, $z=0$ for our study), so that the synchrotron radiation overcomes another main source of energy loss due to inverse Compton scattering with the ambient CMB photons. We also mention that the electron energy loss due to the synchrotron radiation is proportional to the square of electron energy, as well as the magnetic field squared, and hence is suppressed for the lighter WIMP mass. For instance, the energy loss due to Coulomb interactions with thermal plasma can dominate those due to synchrotron radiation and inverse Compton scattering for the sub-GeV electron energy.  We focus on  the typical weak scale WIMP mass range above GeV scale in this paper. Another reason for our restricting our discussions to the above GeV WIMP mass as well as the solar mass range for the PBH mass is to assume that the kinetic energy of WIMPs is negligible compared with its potential energy under the influence of PBH's gravity. The WIMP halo formation around the PBH can well be disturbed by WIMP kinetic energy leading to a less dense profile. Consequently the WIMP annihilation signals become smaller and the bounds on the allowed PBH DM fraction are expected to be significantly weakened when the kinetic energy is not negligible \cite{Carr:2020mqm,Boudaud:2021irr,Boucenna:2017ghj}. 
For the sub-GeV WIMP mass, on the other hand, the other bounds different from considering the UCMHs around PBHs can give tight bounds such as those due to the enhanced isocurvature perturbations from the Poisson fluctuations in PBH distributions \cite{Afshordi:2003zb,Tashiro:2021kcg,Kadota:2020ahr,Inman:2019wvr,Murgia:2019duy,Oguri:2020ldf,Gong:2017sie,Carr:2018rid,Mena:2019nhm}.
The kinetic energy of WIMP is heavily model dependent \cite{Loeb:2005pm,Bertschinger:2006nq,Gondolo:2012vh, Profumo:2006bv,Gondolo:2016mrz,Green:2003un,Green:2005fa,Bringmann:2006mu} and we leave the study for the scenarios with non-negligible WIMP kinetic energy for future work.

The quantitative comparison with the previous relevant works is in order. The previous papers studying the gamma ray bounds on the PBH-WIMP mixed DM scenarios gave the bounds from the Milky Galaxy of order $f_{PBH}\lesssim {\cal O}(10^{-9})\sim {\cal O}(10^{-7})$ for $m_{\chi}\sim 10^{1}-10^3$ GeV when the primary annihilation channel is $\chi \chi \rightarrow b \bar{b}$ \cite{Scott:2009tu,Gondolo:1999ef,Lacki:2010zf,Boucenna:2017ghj,Adamek:2019gns,Eroshenko:2016yve,Carr:2020mqm,Cai:2020fnq,Delos:2018ueo,Kohri:2014lza,Bertone:2019vsk,Ando:2015qda,Hertzberg:2020kpm,Yang:2020zcu,Zhang:2010cj}. $f_{PBH}\equiv \Omega_{PBH}/\Omega_{DM}$ is the PBH abundance fraction with respect to the total dark matter abundance and the total dark matter consists of PBHs and WIMPs ($\Omega_{DM}=\Omega_{PBH}+\Omega_{\chi}$). There have been recent studies discussing the CMB and 21cm bounds which also demonstrated the bounds comparable to the gamma rays \cite{Tashiro:2021xnj,Yang:2020zcu}.
Our radio bounds lead to the upper bounds $f_{PBH}\lesssim {\cal O}(10^{-8})\sim {\cal O}(10^{-7})$
for $m_{\chi}\sim 10^{1}-10^3$ GeV when $\chi \chi \rightarrow b \bar{b}$, and hence support the claim of the other frequency observations that the PBH DM and WIMP DM cannot coexist.

Section \ref{setup} outlines our model setup where we illustrate the WIMP halo possessing the steep density profile formed around a PBH. Due to the large WIMP density around a PBH, one can expect the large enhancement in the WIMP annihilation which is proportional to the WIMP density squared. Section \ref{results} then presents the expected radio signals due to the synchrotron radiation from the energetic \epm arising from the WIMP annihilation, and compare them with the observation data to put the bounds on the allowed PBH abundance. Sec \ref{discon} is devoted to the discussion/conclusion. Throughout the paper, our discussions assume $f_{PBH}\ll 1$ unless stated otherwise (equivalently $1-f_{PBH}\approx 1$ and this assumption is justified in our quantitative analysis in Section \ref{results}). 


\section{Setup}
\label{setup}
\begin{figure}[!htbp]

     \begin{tabular}{c}

              \includegraphics[width=0.5\textwidth]{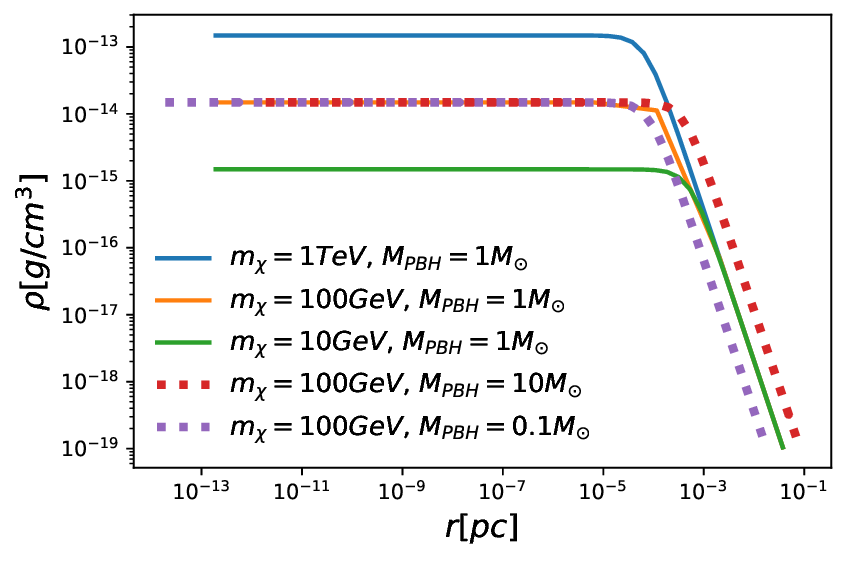} 
    \end{tabular}
       

    
     \caption{The UCMH density profile around a PBH as a function of the radius. The WIMP mass $m_{\chi}$ and the PBH mass $M_{PBH}$ are varied. Each curve covers the radius from $r_{min}=4GM_{PBH}$ to the turn around radius at the matter radiation equality $r_{ta}(z_{eq})$.}
   \label{rhoR}
\end{figure}

We first outline the UCMHs (ultracompact minihalos) around primordial black holes which can posses the steep WIMP density profile $\rho(r)\propto r^{-9/4}$ \cite{Adamek:2019gns,Serpico:2020ehh,Gondolo:1999ef,Boudaud:2021irr,Eroshenko:2016yve, Boucenna:2017ghj}. 
The WIMP particles can accrete to PBHs when the PBHs are formed in the radiation dominated epoch, and we aim to study the effects of the resultant enhanced WIMP annihilation on the radio signals. The WIMP density profile can be estimated via the spherical collapse model where the turn around radius is numerically estimated as \cite{Adamek:2019gns}
\ba
r_{ta}\approx (R_S t_{ta}^2)^{1/3}
\ea
which represents the scale at which a WIMP particle decouples from the background Hubble flow under the gravitational influence of a PBH. $R_S=2GM_{PBH}$ is the Schwarzschild radius, $M_{PBH}$ is the PBH mass, and $t_{ta}$ is the turn around time when a WIMP particle stops moving away from a PBH and starts falling towards it at $r_{ta}$. Assuming each mass shell density matches the background density at the turn around, the WIMP halo density profile during the radiation dominated epoch can be estimated as 
\ba
\rho_{sp}(r)
\approx \frac{\rho_{eq}}{2} \left( \frac{t_{ta}}{t_{eq}}  \right)^{-3/2}
\approx
\left(
\frac{\rho_{eq}}{2}
\right)
t^{3/2}_{eq}
(2GM_{PBH})^{3/4}
r^{-9/4}~.
\ea
Such a steep profile $\rho_{sp}(r) \propto r^{-9/4}$ (possessing a "spike'' in contrast to a more conventional ``cusp" such as in the NFW profile \cite{Navarro:1995iw}) was also verified in the numerical simulations \cite{Adamek:2019gns,Serpico:2020ehh}. While the UCMH around a PBH can grow during the matter domination epoch according to the secondary infall mechanism, we conservatively consider the annihilation only from the region inside the turn around radius at the matter radiation equality $r_{ta}(z_{eq})\sim 0.04 (M_{PBH}/M_{\odot})^{1/3}$ pc. It was verified numerically that such a steep profile is maintained even if the outer part of the halo follows the conventional less steep profile (e.g. NFW profile) \cite{Adamek:2019gns,Serpico:2020ehh}. The spike profile $\rho \propto r^{-9/4}$ makes the WIMP annihilation so efficient around the center of a halo, so that the density can saturate when the WIMP annihilation time scale becomes comparable to the age of the WIMP halo. The density of such an annihilation plateau in the core region can be taken account of as
\ba
\rho_{max} (t) \approx  \frac{m_{\chi}}{\langle \sigma v \rangle (t-t_i) }
\ea
where $m_{\chi}$ is the WIMP mass, $\langle \sigma v \rangle$ is the thermally averaged WIMP annihilation cross section, $t_i$ is the formation time of UCMH (in our quantitative calculations for the signals from the Milky Way Galaxy, $t\gg t_i$ is the age of the Universe and $t-t_i \approx t$). The UCMH density profile hence reads \cite{Gondolo:1999ef,Tashiro:2021xnj,Shelton:2015aqa, Yang:2020zcu,Fields:2014pia,Shelton:2015aqa,Shapiro:2016ypb,Johnson:2019hsm,Adamek:2019gns,Boucenna:2017ghj,Boudaud:2021irr,Carr:2020mqm,Eroshenko:2016yve,Ullio:2001fb}
\ba
\rho_{UCMH}(r) = 
 \begin{cases}
  0     \quad &{\rm for}~ r < 4GM_{PBH}  \\
   \frac{\rho_{sp}(r) \rho_{max}(r)}
       {\rho_{sp}(r)+\rho_{max}(r)}
  \quad &{\rm for}~ 4GM_{PBH} \leq r < r_{\rm ta}(t_{\rm eq}) . \\
 \end{cases}
 \label{rho1b}
 \ea
 A WIMP particle is captured by a black hole for $r<2R_S=4GM_{PBH}$ setting the inner radius of a halo 
 \cite{Vasiliev:2007vh,Sadeghian:2013laa}. The second line comes from integrating $\dot{n}_{\chi}=-n_{\chi}^2 \langle \sigma v \rangle $ with $\rho_{\chi}=n_{\chi} m_{\chi}$. The UCMH profiles for a few representative parameter sets are shown in Fig. \ref{rhoR}.  There is a transition from the inner spike region $\rho_{max}$ forming a core with a flat amplitude to the outer spike region $\rho_{sp}\propto r^{-9/4}$ at a radius $r_{core} \propto m_{\chi}^{-4/9} M_{PBH}^{1/3}$ characterized by $\rho_{max}(r_{core})=\rho_{sp}(r_{core})$. The annihilation plateau amplitude $\rho=m_{\chi} \langle \sigma v \rangle^{-1} t^{-1}$ increases for a bigger $m_{\chi}$ because of the smaller WIMP number density to annihilate and decreases for a bigger annihilation cross section and at a later time because more WIMP particles annihilate.  The majority of annihilation products come from the region $r \sim r_{core}$ and considering the profile at a larger radius $r \gtrsim r_{ta}(z_{eq})$ does not significantly affect our discussions because of the small density \cite{Eroshenko:2016yve,Adamek:2019gns,Boucenna:2017ghj,Carr:2020mqm,Cai:2020fnq,Yang:2020zcu,Tashiro:2021xnj, Kadota:2021jhg}. The annihilation signal contributions from the regions well inside $r_{core}$ does not significantly affect our discussions either because of the small volume despite the large WIMP density. Our goal is to estimate the radio signals from these UCMHs around PBHs in our Milky Way Galaxy. 

\section{Results}
\label{results}
We aim to obtain the radio bounds on the allowed PBH fraction in the PBH-WIMP mixed DM scenarios. The energetic \epm originating from the WIMP annihilation can emit synchrotron radiation in the radio frequency and one can expect the enhancement of such radio signals due to the enhanced WIMP density around PBHs. 

It is customary to use the brightness temperature $T_B[K]$ in the radio observations
($k_B$ is the Boltzmann constant)
\ba
T_B=\frac{c^2}{2\nu^2 k_B}I_{\nu}
\ea
which is related, by the Rayleigh-Jeans law, to the specific intensity $I_{\nu}[$erg cm$^{-2}$s$^{-1}$Hz$^{-1}$sr$^{-1}]$ (the flux density per solid angle per unit frequency).
   For the signal from a direction $(l,b)$ (longitude $l$ and latitude $b$ in the galactic coordinate), the specific intensity is obtained by integrating the emissivity $j_{syn}$ along the line of sight
   \ba
   I_{\nu} (l,b)=\int_{LOS} ds \frac{j_{syn}(\nu,r(s,l,b))}{4\pi}
   \ea
    with the emissivity at a frequency $\nu$
 \ba
 j_{syn}(\nu,r)=2\int^{M_{DM}}_{m_e} dE \frac{d n_e(E,r)}{dE} P_{syn}(\nu, E, r)
 \ea
 where $P_{syn}$ is the synchrotron power emitted by an electron at a frequency $\nu$ averaged over all directions \cite{2011hea..book.....L}. A factor 2 takes account of the contributions from both electrons and positrons. For a given point with the distance $s$ along the line of sight, the distance $r$ from the Galactic center can be given in terms of the galactic coordinate $r^2(s,l,b)= s^2- 2 s r_{\odot}\cos b \cos l +r_{\odot}^2$ (where $r_{\odot}=8.33$ kpc is the Sun's distance from the Galactic center).
To estimate the radio signals of our interest, we need to know the electron/positron energy spectrum originating from the WIMP annihilation.
Those energetic electrons/positrons diffuse through the galactic medium while losing energy due to the processes such as the synchrotron radiation and the inverse Compton scattering. The final electron number density per unit energy $d n_e(E,r)/dE$ can be obtained by solving the diffusion-loss equation \cite{2011hea..book.....L,Baltz:1998xv,Delahaye:2007fr}
\ba
\frac{\partial }{\partial t} \frac{dn_e(E,r)}{dE}=
\nabla\cdot 
\left[
  K(E,r) \nabla \frac{dn_e(E,r)}{dE}
  \right]
+
\frac{\partial}{ \partial E}
\left[
  b(E,r)
  \frac{dn_e(E,r)}{dE}
  \right]
+
Q(E,r)
\ea
where $E$ denotes the energy of electrons. The source term in our PBH-WIMP mixed DM scenario is 
\ba
Q(E,r)=n_{PBH} 
 \Gamma_{PBH}^f \frac{dN^f_e(E,r)}{dE}
\label{qsource}
\ea
\ba
n_{PBH}=\frac{f_{PBH}\rho_{Gal}}{M_{PBH}}, ~\Gamma_{PBH}^f=\frac{1}{2m_{\chi}^2} \int dR 4 \pi R^2 \rho_{UCMH}^2 \langle \sigma v \rangle_f ~.
\ea
We for concreteness assume Majorana particles for WIMPs. $K, b$ represent respectively the diffusion coefficient and energy loss coefficient \cite{2011hea..book.....L,Colafrancesco:2005ji,Hooper:2012jc,Storm:2016bfw,McDaniel:2017ppt,Bertone:2008xr,Fornengo:2011iq,Buch:2015iya,Cirelli:2013mqa}. Morphologically the WIMP annihilation signals from the unresolved UCMHs around PBHs can be interpreted as the signals from the `decaying' PBHs dressed by WIMPs. Hence the source term resembles that for a decaying DM scenario with the decay rate $\Gamma$. $dN_e^f/dE$ represents the electron energy distribution per one WIMP annihilation with a given final state channel $f$ (we will later consider $\chi \chi \rightarrow b \bar{b}$ channel and $\chi \chi \rightarrow e^+ e^-$ channel for illustration purposes).
We assume the PBH number density distribution $n_{PBH}(r)$ follows the underlying Galaxy density profile $\rho_{Gal}(r)$ modeled by the NFW profile.
In estimating the synchrotron radio emission from the dressed PBHs, we simply treat the UCMHs as unresolved point-like objects which are embedded in the Milky Way Galaxy's magnetic fields \footnote{The length scale for the \epm to lose a half of its energy and also its diffusion length scale are larger than the scales relevant for the UCMHs \cite{Saito:2010ts,Colafrancesco:2006he}.}. The tidal disruptions of substructures inside a host halo are still under active debate. While the outskirts of the compact halos can be tidally stripped away, the dense core parts may well survive where the dominant DM annihilation signals come from \cite{Goerdt:2006hp,vandenBosch:2017ynq,2015MNRAS.454.1697X,Berezinsky:2007qu,Delos:2019lik,Green:2019zkz,2010PhRvD..81j3529B,Green:2006hh,Goerdt:2006hp,Schneider:2010jr}. We simply assume the core of an UCMH is unaffected by tidal forces and leave the further studies including the effects of tidal disruptions and the detailed UCMH spatial distributions for future work. 
We accordingly consider the annihilation only from the core region of UCMH $r \leq r_{ta}(z_{eq})\sim 0.04 (M_{PBH}/M_{\odot})^{1/3}$ pc in calculating $\Gamma^f_{PBH}$. One can obtain the steady state equilibrium solution for $d n_e/dE$ by the Green's function method \cite{Colafrancesco:2005ji,Baltz:1998xv,Cirelli:2010xx,Buch:2015iya,McDaniel:2017ppt} and we use the PPPC4DM package for our numerical calculations \cite{Cirelli:2010xx,Buch:2015iya}.
To estimate the signals, one needs to specify the cosmic ray's diffusion model and our choices are described in the following. In treating the charged particle propagation, we adopt the conventional parameterization of the diffusion coefficient $K=K_0 (E/GeV)^{\delta}$ which is position-independent but depends on the energy. Those diffusion coefficients can be determined from the cosmic ray abundance measurements such as those of boron-to-carbon ratio and the conventional propagation parameter sets adopted in our analysis are $(\delta, K_0[$kpc$^2/$Myr$],L[$kpc]) =(0.55,0.00595,1), (0.70,0.0112,4),(0.46, 0.0765,15), often referred to respectively as MIN, MED, MAX models (the primary anti-proton fluxes due to the DM annihilation in the Milky Way halo are minimal, median or maximal at the solar position) \cite{Donato:2003xg,Delahaye:2007fr,Cirelli:2010xx,Buch:2015iya,Blum:2010nx}. The cylindrical diffusion zone has the hight $2L$ and radius $20$ kpc. The MED propagation parameters are commonly used in the literature and we present our results using the MED model in this section. Using MIN and MAX models do not affect our discussions as demonstrated in the discussion section.  $b(E)=b_{syn}+b_{ICS}+b_{Coul}+b_{brem}$ includes the effects of energy loss due to synchrotron radiation, inverse Compton scattering, Coulomb interaction and bremsstrahlung \cite{Buch:2015iya,2011hea..book.....L,Colafrancesco:2005ji,Hooper:2012jc,Storm:2016bfw,McDaniel:2017ppt,Bertone:2008xr,Fornengo:2011iq,Cirelli:2013mqa}. We modeled the energy loss rate $b(E)$ as in Ref. \cite{Buch:2015iya}. 
We note that, for the parameter range of our interest (the typical energy scale of \epm is $\gtrsim $ GeV), the energy loss is dominated by the synchrotron radiation and the inverse Compton scattering. The estimation of energy loss due to the synchrotron radiation
$b_{syn}(E)\propto E^2 B^2$ requires us to specify the magnetic field profile $B$ for which we adopt the following conventional profile \cite{Strong:1998fr,Cirelli:2010xx}
\ba
B(r,z)=B_0 \exp[
-(r-r_{\odot})/r_B - |z|/z_B
]
\ea
The common choices for the magnetic field configuration parameters are $B_0=4.78$ $\mu$G, $r_B=10$ kpc, $z_B=2$ kpc referred to as the model MF1 and $(B_0 [\mu$G], $r_B$ [kpc], $z_B$ [kpc])=$(5.1,8.5,1)$ and $(9.5, 30,4)$ respectively referred to as the MF2 and MF3 models. MF1 is the configuration similar to that used in GALPROP code \cite{Strong:1998fr}, MF2 is based on Ref \cite{Sun:2007mx} which has a bigger spatial gradient with a higher magnitude at the Galactic center and MF3 has a much bigger magnitude at the Earth location and the magnetic field region extends much further than the other models \cite{Strong:2011wd}. The effects due to the variation in the magnetic field parameters are degenerate with those due to the variation of the aforementioned propagation parameters. In fact, we found the difference in the synchrotron radiation signals among these three magnetic field parameter sets is much smaller than that among aforementioned MIN/MED/MAX models. We therefore simply present our findings using the MF1 model for the magnetic field configuration. 
The estimation of energy loss due to the inverse Compton scattering requires the photon distributions which \epm scatters with, and, following Ref. \cite{Cirelli:2010xx}, we adopt the blackbody spectrum for the CMB photons and the map from GALPROP for infrared light and star light \cite{Vladimirov:2010aq}. We note, at radio frequencies, the energy loss is dominated by the synchrotron radiation in the Galactic center for the parameter range of our interest due to the large magnetic field where the dominant signals come from (in the presence of the reasonably strong magnetic field bigger than that corresponding to the equivalent CMB energy density $B \gtrsim 3.25(1+z)^2 \mu $G (the redshift $z=0$ in our calculations)).

\begin{figure}[!htbp]

     \begin{tabular}{cc}
              \includegraphics[width=0.5\textwidth]{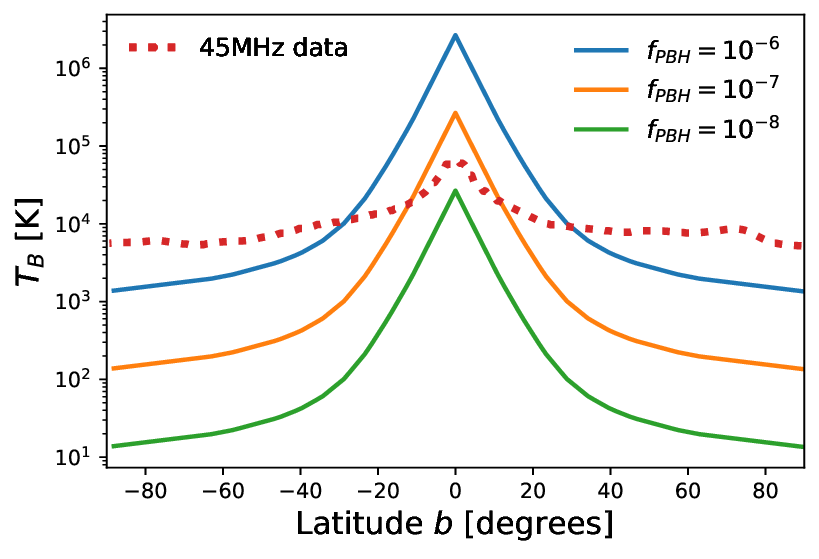}
       &
                       \includegraphics[width=0.5\textwidth]{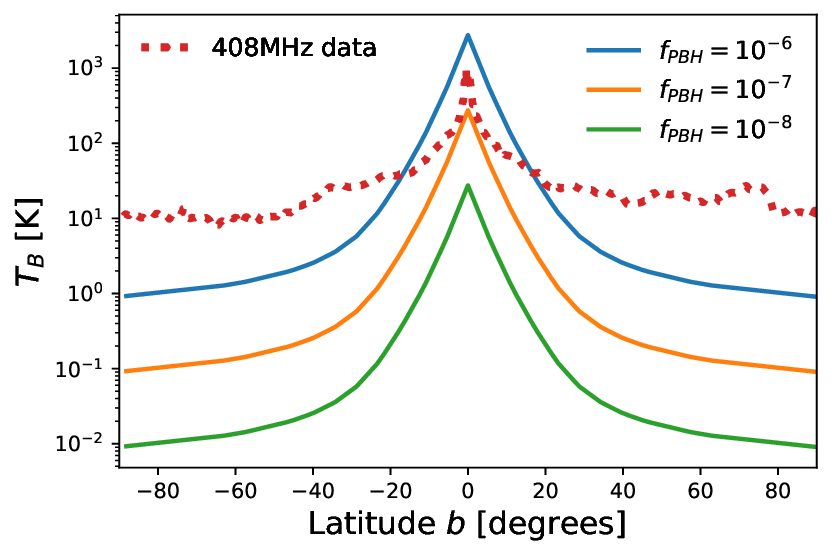}
    \end{tabular}
    
         \caption{
           The brightness temperature of the emitted synchrotron radiation signals at the frequencies 45 MHz and 408 MHz as a function of the galactic latitude $[-90^{\circ},90^{\circ}]$. The galactic longitude is set to $l=0^{\circ}$. The WIMP primary annihilation channel is $\chi \chi \rightarrow b \bar{b}$, $M_{PBH}=1M_{\odot}$. The PBH DM fraction $f_{PBH}$ is varied for illustration. The observed radio emission data are also shown for comparison. 
         }


   \label{bvsTb}
\end{figure}

Using these model parameters, we can make predictions on the expected signals in our PBH-WIMP mixed DM scenarios as shown in Fig. \ref{bvsTb} along with the observed radio emission data. The figures are plotted as a function of the galactic latitude $b$ along the galactic longitude $l=0$. The chosen parameter sets are $m_{\chi}=100$ GeV, $\langle \sigma v \rangle =3 \times 10^{-26} cm^3/s, M_{PBH}=1M_{\odot}$ and the primary annihilation channel $\chi \chi \rightarrow b \bar{b}$. The MED propagation model, MF1 magnetic field model and the NFW density profile are used for the Milky Way Galaxy. The spherical symmetry is assumed in the signal calculations. The signals and foregrounds are bigger towards the Galactic center because of the bigger magnetic field and DM density.
While we chose the data at 45 MHz and 408 MHz because of the data availability, some further comments on our choice of frequencies are in order. 
For the WIMP mass below 10 TeV, the synchrotron radiation signals in the Galaxy typically peak at below tens of GHz with a magnetic field of order $\mu G$.
In choosing the observation data to compare with the signals, we also prefer those below a GHz to ensure that the measured temperature is dominated by the synchrotron radiation to avoid being plagued by the other contributions such as the free-free, thermal dust and CMB radiation. We accordingly choose the observation maps with a large sky coverage at 45 MHz and 408 MHz from Ref. \cite{Guzman:2010da,pllegacy,Fornengo:2014mna,Planck:2016gdp} (for which the data are available for the whole sky). The angular resolution of 45 MHz map is worse than that of 408 MHz map (the survey angular resolutions are respectively 5 degrees and 0.85 degrees) and the flux in the Galactic central region in our plots is more smoothed out in the former. These maps suffice for our purpose to put the bound on the PBH fraction.
We constrain $f_{PBH}$ by demanding that the expected DM signals should not exceed the observed data with 3 $\sigma$, $T_{DM}<T_{data}+3\sigma_{data}$ at all latitudes along the longitude $l=0$. The map at 45 MHz  was obtained by combining the southern and northern surveys and the root mean square temperature noise is 300 K/2300 K for the southern/northern data and that at 408 MHz is of order one Kelvin. The error bars of the radio data used in our analysis hence are an order of magnitude smaller than the observed temperature and do not give appreciable impacts on our bounds, especially in comparison to the systematic uncertainties in theoretical cosmic ray propagation modeling \cite{Fornengo:2014mna,Buch:2015iya,Cirelli:2016mrc}. We can see from these figures that the bounds mainly come from the central region of the sky. For the region outside the central Galactic region where the magnetic field is smaller, the synchrotron signal typically peaks at a frequency below one MHz and the majority of signals can lie below the observable frequencies accessible by the radio telescopes. The tight bounds hence can come from the central region of our Galaxy even though the synchrotron radiation can cover a wide range of frequency beyond its peak frequency. We will show in the discussion section that the bound would not change by more than an order of magnitude even if we take account of the large systematic uncertainties such as those in the propagation models and the magnetic field profile.
\begin{figure}

     \begin{tabular}{cc}
              \includegraphics[width=0.5\textwidth]{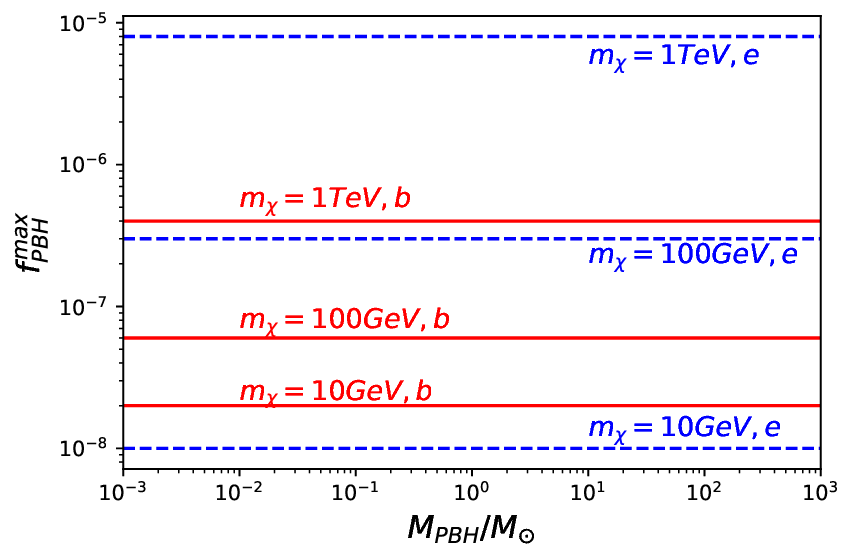}
       &
                         \includegraphics[width=0.5\textwidth]{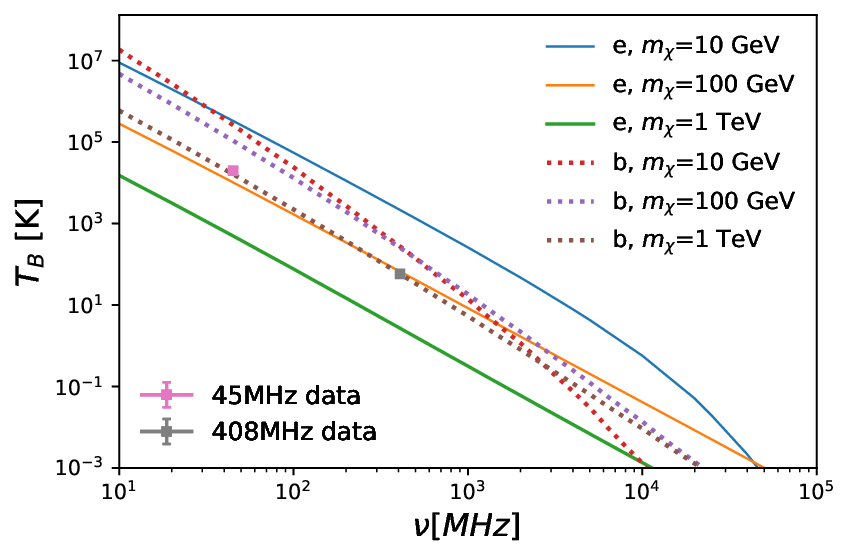}
       
    \end{tabular}
       
    
     \caption{
       [Left] The upper bounds for the allowed PBH DM fraction. The chosen WIMP annihilation channels are $\chi \chi \rightarrow b \bar{b}$ and $\chi \chi \rightarrow e^+ e^-$ (e.g. $m_{\chi}=100$ GeV$,b$ means it is for $\chi \chi \rightarrow b \bar{b}$ with the WIMP mass 100 GeV). 
       [Right]  The brightness temperature of the synchrotron radiation as a function of the frequency at the Galactic center ($(l,b)=(0^{\circ},0^{\circ})$). The WIMP primary annihilation channels of $\chi \chi \rightarrow b \bar{b}$ and $\chi \chi \rightarrow e^+ e^-$ are shown for comparison. 
       $f_{PBH}=10^{-7}$ is chosen for illustrative purposes. The error bars for the observation data points are too small to be seen in this figure.}
     

   \label{upperfpbh}
\end{figure}
The left panel of Fig. \ref{upperfpbh} shows the consequent upper bounds on $f_{PBH}$ for $\chi \chi \rightarrow b \bar{b}$ illustrated in Fig. \ref{bvsTb}. We can see that the PBH DM abundance is negligible ($f_{PBH}\ll 1$) when the WIMP DM is the rest of DM of our Universe. One can do the same exercises for the leptonic channel too and the bounds for the $\chi \chi \rightarrow e^+ e^-$ annihilation scenario are also shown in Fig. \ref{upperfpbh}. Note our bounds on $f_{PBH}$ is insensitive to the values of $M_{PBH}$, which is also the case for the gamma ray bounds on $f_{PBH}$ \cite{Scott:2009tu,Gondolo:1999ef,Lacki:2010zf,Boucenna:2017ghj,Adamek:2019gns,Eroshenko:2016yve,Carr:2020mqm,Cai:2020fnq,Delos:2018ueo,Kohri:2014lza,Bertone:2019vsk,Ando:2015qda,Hertzberg:2020kpm,Yang:2020zcu,Zhang:2010cj}. This is simply because of the cancellation of the $M_{PBH}$ dependence between $\Gamma_{PBH} \propto M_{PBH}$ and $n_{PBH} \propto M_{PBH}^{-1}$. 
The dependence of the upper bounds of $f_{PBH}$ on $M_{PBH}$ however shows up, for instance, when the annihilation cross section is velocity dependent \cite{Kadota:2021jhg} or when the kinetic energy of WIMP is not negligible preventing the steep profile around a PBH (which can be the case for $M_{PBH}\lesssim 10^{-3} M_{\odot}$ if $m_{\chi}\lesssim 10 $ GeV \cite{Adamek:2019gns,Eroshenko:2016yve,Kadota:2020ahr,Carr:2020mqm, Boudaud:2021irr,Boucenna:2017ghj}). 
The non-trivial relative ordering of the bounds illustrated in this figure stems from the non-trivial frequency dependence of the signals and the observational data used. The relative comparison of the signals for the $b \bar{b}$ and $e^+ e^-$ annihilation channels as a function of the frequency is illustrated in the right panel of Fig. \ref{upperfpbh}. The hadronic channel $\chi \chi \rightarrow b \bar{b}$ tends to have softer \epm energy spectrum than the leptonic channel $\chi \chi \rightarrow e^+ e^-$ due to the hadronization. Consequently, $b \bar{b}$ annihilation scenario is more tightly constrained by the 45 MHz data than by 408 MHz data while the upper bound on $f_{PBH}$ for $e^+ e^-$ annihilation scenario comes from 408 MHz data rather than from 45 MHz data for the electroweak scale WIMP mass range of our interest.
Our synchrotron radio emission bounds turned out to be an order of magnitude less tight than those from the gamma ray and CMB bounds \cite{Boucenna:2017ghj,Adamek:2019gns,Eroshenko:2016yve,Carr:2020mqm,Yang:2020zcu,Tashiro:2021xnj,Yang:2020zcu}. Our synchrotron radiation bounds nevertheless give the independent support for the claim that PBH DM and WIMP DM cannot coexist. We limited our analysis to 45 MHz and 408 MHz data sets for their large sky coverage and we found the bounds on $f_{PBH}$ can change by a factor a few between these data sets for a given annihilation channel. Further improvement on the bounds could well be achievable by using the other data sets at the frequencies optimized to the specific DM annihilation models. The more systematic studies using larger radio observation data sets and more concrete particle physics models are left for future work. 

\begin{figure}[!htbp]

     \begin{tabular}{c}
              \includegraphics[width=0.5\textwidth]{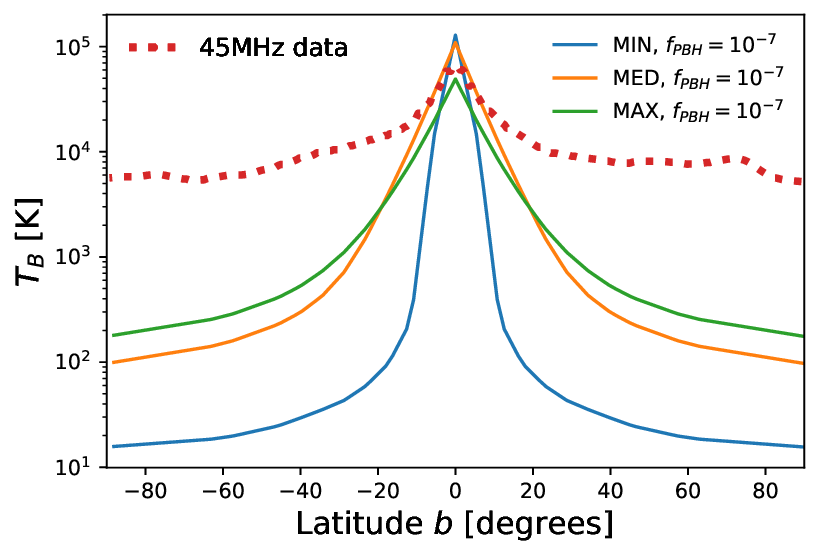}

    \end{tabular}
       
    
     \caption{
       The brightness temperature as a function of the galactic latitude $[-90^{\circ},90^{\circ}]$ at 45 MHz. The galactic longitude is set to $l=0^{\circ}$. The WIMP primary annihilation channel is $\chi \chi \rightarrow b \bar{b}$, $M_{PBH}=1M_{\odot}, m_{\chi}=100$ GeV. The cosmic ray propagation parameters are varied among the MIN, MED, MAX models for comparison. $f_{PBH}=10^{-7}$ is chosen for illustrative purposes.}

   \label{minmax}
\end{figure}

\section{Discussion/Conclusion}
\label{discon}

The synchrotron radiation predictions involve theoretical uncertainties such as those in the cosmic ray propagation model and magnetic field profile in Galaxy.
Before concluding our discussions, let us briefly discuss how the predicted signals can be affected by these model uncertainties. The widely used propagation models in the literature are so-called MIN, MED, MAX models as mentioned in the last section, and we adopted the commonly used MED model in the last section.
Fig. \ref{minmax} shows the signals using the MIN and MAX model parameters for comparison.
 The radiation from the MIN model is confined to the Galactic center region while the MAX model has the thicker diffusive halo and the radiating \epm propagation extends to a larger region. 
 While the signal predictions are indeed affected by the theoretical model uncertainties, the changes are within an order of magnitude among these conventional models and our bounds on $f_{PBH}$ consequently would not change by more than an order of magnitude.
We also varied the magnetic field profiles among MF1, MF2 and MF3 mentioned in the last section, and the difference in the brightness temperature is at most a factor a few and they give smaller effects on $T_B$ than the propagation model variations shown in this figure. 
Hence we conclude that, even though the quantitative discussions can change due to the systematics in the theoretical modeling of cosmic ray propagation, the non-coexistence of PBH DM and WIMP DM is robust.

As commonly discussed in the literature discussing the UCMHs around PBHs, our tight bounds depend on the assumption that the kinetic energy of WIMPs is negligible in their forming the halos around PBHs. Accordingly we limited the values of $M_{PBH}, m_{\chi}$ to the range where the WIMP kinetic energy is of order a percent or less compared with the potential energy \cite{Adamek:2019gns,Eroshenko:2016yve,Kadota:2020ahr,Carr:2020mqm, Boudaud:2021irr,Boucenna:2017ghj}. If the kinetic energy can interfere the gravitational influence of PBH on the WIMPs, the WIMP halos around PBHs can be less steep and hence the consequent annihilation signals will be reduced \cite{Carr:2020mqm,Boudaud:2021irr,Boucenna:2017ghj}. Such scenarios where the kinetic energy cannot be negligible is heavily dependent on the particle physics model assumptions such as the nature of WIMP kinetic decoupling from the thermal plasma in the early Universe, and such scenarios in view of the radio signals are left for future work. We focused on the signals from the Milky Way Galaxy in this paper and we plan to study the extra-galactic signals including the future prospects from the forthcoming radio experiments such as the SKA in our future work.
\\

KK thanks P. Gondolo for the useful discussions. This work was partially supported by Grants-in-Aid for Scientific Research from JSPS (21K03533, 21H05459).



\bibliography{/Users/kenji/GD/work/paper/kenjireference}



\end{document}